\documentstyle[aps,preprint,epsf]{revtex}
\begin{document}
\draft

\title{Quantum critical phenomena
of long-range interacting bosons\\
in a time-dependent random potential}
\author{Kihong Kim\cite{kk}}
\address{Department of Physics, Ajou University, Suwon 442-749, Korea}
\maketitle 
\begin{abstract}
We study the superfluid-insulator transition
of a particle-hole symmetric system of 
long-range interacting bosons in a time-dependent
random potential in two dimensions, using the momentum-shell
renormalization-group method.
We find a new stable fixed point with non-zero values
of the parameters representing the short- and long-range
interactions and disorder when the interaction is 
asymptotically logarithmic.
This is contrasted to the non-random case with a logarithmic
interaction, where the transition is argued to be first-order,
and to the $1/r$
Coulomb interaction case, where either a first-order transition
or an XY-like transition is possible depending
on the parameters. We propose that our model may be relevant in studying
the vortex liquid-vortex glass transition of interacting vortex lines
in point-disordered type-II superconductors.
\\ 

\end{abstract}

\narrowtext
In recent years, there has been great interest
in quantum critical phenomena occurring in
systems of interacting bosons \cite{fg,fwgf,wk,fgg,fis,kw,cfgwy,sgcs}.
This has largely been motivated by several beautiful
experiments on the superconductor to insulator
transition in thin metallic films at very low temperature 
\cite{hlg,hp,lhng,yk,mcg,mk}.
In many models, the superconductor (or charged superfluid)
 to insulator 
(or Bose glass) transition in disordered films 
is considered as a localization transition
of bosonic Cooper pairs in an external random potential 
\cite{fg,fgg,fis,kw,cfgwy}.
In non-random systems, the superfluid to (Mott) insulator 
transition can be triggered by changing
the density of bosons through critical values commensurate
with the underlying periodic lattice \cite{fg,fwgf}.
Since Cooper pairs are charged, one needs to consider 
the influence of long-range interactions on the nature
of the critical phenomena. In the absence of disorder,
the superconductor-Mott insulator transition of charged
bosons in two dimensions interacting 
by the $1/r$ Coulomb potential has been studied 
by Fisher and Grinstein via renormalization-group methods \cite{fg}.
They showed that when the (bare) Coulomb
interaction is sufficiently weak, 
the transition belongs to the three-dimensional
XY-model universality class, whereas when the Coulomb 
interaction is sufficiently strong, 
it is first-order. In this paper, we generalize Fisher and 
Grinstein's model to the cases with 
the $1/r$ or logarithmic interaction in
a {\it time-dependent} random potential. Time dependence
of the potential in 
classical static critical phenomena is unimportant \cite{sgcs}. 
In quantum critical phenomena, however,
it can lead to new universality classes as will be 
shown in the present work. 
As well as being relevant in studying novel
quantum critical phenomena, our model may be also 
a useful representation of a system of long-range
interacting vortex lines in {\it point-disordered}
type-II superconductors in an external magnetic field.
According to the mapping originally due to Nelson,
vortex lines in three spatial 
dimensions are considered
as the world lines of bosons residing 
in two spatial dimensions perpendicular to the magnetic field 
\cite{nel,fl,ln}.
The dimension parallel to the magnetic field is mapped to the
imaginary time dimension of a bosonic system. 
Therefore the static random 
potential provided by isotropically distributed
impurities in superconductors can be regarded as a 
time-dependent potential that 
is random in both space and time.
The quantum superconductor-insulator transition 
of interacting bosons is analogous to
the classical vortex liquid-vortex
glass transition of interacting vortex lines. 
We caution that in our model,
the long-range interaction between vortex lines is confined to the
plane perpendicular to the magnetic field. 
Applying the standard momentum-shell 
renormalization-group method 
to the {\it logarithmically} interacting case, 
we discover 
a new stable fixed point in two dimensions
with non-zero values in the short- and long-range interactions and
disorder when there exists a time-dependent random potential, whereas
we find a first-order transition in the non-random case.
Since the interaction between vortex lines is logarithmic 
in a wide range of length scales when the magnetic field
is close to the upper critical field \cite{hr},
we propose a possibility that our new random fixed point 
may describe the real vortex liquid-vortex
glass transition in 
point-disordered type-II superconductors.
On the other hand, when the interaction is the $1/r$ Coulomb interaction,
we find a behavior analogous to Fisher and Grinstein's results in a non-random
system.

We consider the $(d+1)$-dimensional classical action
for the $m$-component classical complex Bose field
$\psi_i(\vec{x},\tau)$ coupled to the $scalar$ gauge field
$A(\vec{x},\tau)$
mediating the long-range interaction between charge $e$ bosons
in $d$ spatial and one temporal dimensions: 
\begin{eqnarray}
S&=&\int d^dx\int d\tau \sum_{i=1}^m \left[
a\vert \left( \partial_\tau 
-ieA\right)\psi_i\vert^2
+b\vert \nabla\psi_i\vert^2 \right.\nonumber\\
&&+\left. r(\vec{x},\tau)\vert\psi_i\vert^2
\right]  
+\int d^dx\int d\tau\sum_{i,j=1}^mu
\vert\psi_i\vert^2\vert\psi_j\vert^2\nonumber\\
&+&{1 \over 2}\sum_{\vec{k},\omega}(ck^\sigma+gk^2)
\vert A(\vec{k},\omega)\vert^2,
\label{eq:action}
\end{eqnarray}
where $A(\vec{k},\omega)$ is the Fourier transform of
$A(\vec{x},\tau)$.
$a$, $b$, $u$, $c$ and $g$ are positive constants and
$r(\vec{x},\tau)$ is a Gaussian random function satisfying
\begin{eqnarray}
&&r(\vec{x},\tau)=r_0+\delta r(\vec{x},\tau),~~~
\langle\delta r(\vec{x},\tau)\rangle=0,\nonumber\\
&&\langle\delta r(\vec{x},\tau)
\delta r({\vec{x}}^\prime,\tau^\prime)\rangle
=2v\delta(
\vec{x}-{\vec{x}}^\prime)\delta(\tau-\tau^\prime),
\label{eq:random}
\end{eqnarray}
where $\langle\cdots\rangle$ denotes the statistical
averaging and $v$ ($>0$) measures the strength of randomness.
The action $S$ applies to the cases where 
there is a particle-hole symmetry and the interaction
potential between bosons is proportional to $1/r^{d-\sigma}$
in the long distance limit,
$r$ being the distance between bosons. 
The $1/r$ Coulomb interaction
corresponds to $\sigma=d-1$, 
while the {\it asymptotically} logarithmic interaction
corresponds to $\sigma=d$.
In the absence of the particle-hole symmetry, one has to replace
the term proportional to $a$ in (\ref{eq:action}) by a term
of the form ${\psi_i}^*(\partial_\tau-ieA)\psi_i$ \cite{fwgf,wk}.
In general, disordered systems do not possess the 
particle-hole symmetry. We note, however, 
that it has been argued recently, in a rather compelling manner,
that the particle-hole asymmetric term is irrelevant 
in the renormalization group sense
and the exact (statistical) particle-hole symmetry is restored
after averaging over disorder 
at the critical points of various kinds of
disorder-driven superfluid transitions,
including the vortex glass transition \cite{fish,dor,muk,pz}.
We accept these arguments and assume that the particle-hole
symmetric action, Eq.~(\ref{eq:action}), 
describes the critical fixed point of 
the vortex glass transition.
When $r$ is non-random and $g$ is zero, the superconductor to
Mott insulator transition described by this model was
studied by Fisher and Grinstein \cite{fg}. In the present paper,
we generalize their work to the cases where 
$r$ is a Gaussian random function of both space and time.

We use the standard replica trick to reduce 
$S$ to a non-random effective action: 
\begin{eqnarray}
S_{\rm eff}&=&\int d^dx\int d\tau \sum_{\alpha=1}^p \sum_{i=1}^m \left[
a\vert \left( \partial_\tau
 -ieA_\alpha\right)\psi_{\alpha i}\vert^2\right.\nonumber\\
&&+\left. b\vert \nabla\psi_{\alpha i}\vert^2
+r_0\vert\psi_{\alpha i}\vert^2
\right]\nonumber\\  
&+& \int d^dx \int d\tau\sum_{\alpha=1}^p\sum_{i,j=1}^mu
\vert\psi_{\alpha i}\vert^2\vert\psi_{\alpha j}\vert^2
\nonumber\\
&-&\int d^d x\int d\tau\sum_{\alpha,\gamma=1}^p\sum_{i,j=1}^mv
\vert\psi_{\alpha i}\vert^2\vert\psi_{\gamma j}\vert^2
\nonumber\\
&+&{1 \over 2}\sum_{\alpha=1}^p\sum_{\vec{k},\omega}(ck^\sigma+gk^2)
\vert A_\alpha(\vec{k},\omega)\vert^2,
\label{eq:replica}
\end{eqnarray}
where $\alpha$ and $\gamma$ are replica indices and the number of replica
$p$ is sent to zero at the end of the calculation. 
In order to apply the momentum shell renormalization group method,
we perform a Fourier transform of Eq.~(\ref{eq:replica}). 
The Fourier-transformed 
action has five types of interaction vertices.
We perform a perturbation expansion in terms of these vertices
to one loop order and integrate out the components with 
wave number $\vec{k}$ in a shell $k_\Lambda/e^l<k<k_\Lambda$,
where $k_\Lambda$ is the momentum space cutoff and $e^l$ is
the rescaling parameter. We recover the original form 
of the action by renormalizing $r_0$, $a$, $b$, $g$, $u$ and $v$ using the
results of the momentum space integration and rescaling frequency, momentum 
and the fields with the rules
\begin{eqnarray}
&&\vec{k}\rightarrow\vec{k}/e^l,~~~
\omega\rightarrow\omega/e^{zl},
\nonumber\\
&&\psi_{\alpha i}(\vec{k},\omega)\rightarrow\psi_{\alpha i}
(\vec{k},\omega) e^{(\zeta_\psi+{d\over 2}+{z\over 2}) l},
\nonumber\\
&&A_\alpha(\vec{k},\omega)\rightarrow A_\alpha
(\vec{k},\omega)e^{(\zeta_A+{d\over 2}+{z\over 2}) l},
\end{eqnarray}
where $z$ is the dynamical exponent and $\zeta_\psi$ and $\zeta_A$ are
rescaling factors of the fields.

We perform our calculations in an $\epsilon$ ($=3-d$) expansion about the
upper critical dimension $d=3$. 
Firstly, we consider the competition of the $k^\sigma$ and $k^2$ terms
in the gauge field propagator. A naive power counting suggests
that the $k^2$ term is irrelevant 
as long as the $k^\sigma$ term with $0<\sigma<2$ is present.
This conclusion is not always true, however, since the $k^2$ term is
renormalized in a perturbation theory, whereas the {\it nonanalytic}
$k^\sigma$ term is not,
and in some cases,
this renormalization
makes the $k^2$ term more relevant than the $k^\sigma$ term.
To confirm this, we fix $a=b=g=1$ and perform a perturbation calculation
assuming that $c$ is irrelevant.
To one loop order, we obtain the exponents
\begin{equation}
z=1-{\tilde{w}\over 3},~~~
2\zeta_\psi=1-d+{\tilde{w}\over 6}
\label{eq:exponent}
\end{equation}
and the recursion relations for $c$:
\begin{equation}
\frac{dc}{dl}=\left(2-\sigma-\frac{m}{12}\tilde w\right)c
\label{eq:g}
\end{equation}
and for $r_0$, $u$ and $v$:
\begin{eqnarray}
\frac{dr_0}{dl}&=&\left(2-\frac{{\tilde w}}{6}\right)r_0
+\frac{4(m+1){\tilde u}}{\sqrt{1+r_0}}
-\frac{4{\tilde v}}{\sqrt{1+r_0}}\nonumber\\
&&+\frac{{\tilde w}\sqrt{1+r_0}}{2},
\nonumber\\
{{d\tilde {u}}\over{dl}}&=&\epsilon \tilde{u}-2(m+4)\tilde{u}^2
+12\tilde{u}\tilde{v}+{{\tilde{u}\tilde{w}}\over 2}-{\tilde{w}^2\over 16},
\nonumber\\
{{d\tilde{v}}\over{dl}}&=&\epsilon \tilde{v}+8\tilde{v}^2
-4(m+1)\tilde{u}\tilde{v}
+{{\tilde{v}\tilde{w}}\over 2},
\label{eq:uv}
\end{eqnarray}
where $\tilde{u}=u\kappa_d/4$, $\tilde{v}=v\kappa_d/4$,
$\tilde{w}=w\kappa_d=e^2\kappa_d$ and $\kappa_d=2
/(4\pi)^{d/2}\Gamma(d/2)$.
The momentum cutoff $k_\Lambda$ was chosen to be equal to 1. 
The recursion relation for the parameter $w\equiv e^2$ is obtained from
the condition that the action is invariant under the gauge transformation
\begin{equation}
\psi=\tilde\psi e^{ie\lambda},~~~A=\tilde A +\frac{d\lambda}{d\tau}
\label{eq:gauge}
\end{equation}
for any function $\lambda(\tau)$ to all orders. We find
\begin{equation}
{{d\tilde{w}}\over{dl}}=(2\zeta_A+2z)\tilde{w}
=\epsilon\tilde{w}-\left(\frac{m+4}{12}\right)\tilde{w}^2,
\label{eq:charge1}
\end{equation}
which possesses a stable fixed point $\tilde w^*=12\epsilon/(m+4)$.
Substituting this value into Eq.~(\ref{eq:g}), we observe 
that the parameter $c$ is indeed irrelevant 
at this fixed point if
$2>\sigma>\sigma_c =2-m\tilde w^*/12=2-m\epsilon /(m+4)$.
In the cases with $0<\sigma<\sigma_c$, where $g$ is irrelevant,
we fix $a=b=c=1$ and $g=0$ and perform a one-loop calculation. It turns out that
the recursion relations (\ref{eq:uv}) remain the same, whereas 
Eq.~(\ref{eq:charge1}) 
is replaced by
\begin{equation}
{{d\tilde{w}}\over{dl}}
=(\epsilon-\epsilon_\sigma)\tilde{w}-{{\tilde{w}^2}\over 3},
\label{eq:charge2}
\end{equation}
where we have introduced another small parameter $\epsilon_\sigma=2-\sigma$.

We consider the $1/r$ Coulomb interaction case first. According to the
results obtained above, the $k^2$ term in the gauge field propagator
is irrelevant to the first order in $\epsilon$ in the physically interesting
case with $\epsilon=1$ (that is, $d=2$) and $m=1$. Then 
$\epsilon-\epsilon_\sigma=1-d+\sigma=0$ and the marginally 
irrelevant parameter $\tilde w$
satisfies 
\begin{equation}
{{d\tilde{w}}\over{dl}}
=-{{\tilde{w}^2}\over 3}.
\label{eq:charge3}
\end{equation}
Integrating this equation and the recursion relations for $\tilde u$
and $\tilde v$ numerically
for various initial values of $\tilde u$, $\tilde v$ and
$\tilde w$, we find
the renormalization group flow diagram qualitatively similar to 
that discovered by Fisher and Grinstein.
That is, there is a plane in the parameter space separating the domain
of attraction to the ($4-\epsilon$)-dimensional random 
XY-model fixed point
with positive ${\tilde u}^*$ and ${\tilde v}^*$ and zero ${\tilde w}^*$
\cite{com1} and the domain 
where all points flow to the region with large negative 
$\tilde u$, large positive $\tilde v$ and zero $\tilde w$.
When the physical value of $\tilde w$ is sufficiently small, the 
flows are attracted to the XY fixed point. 
When the initial $\tilde w$ value is large,
$\tilde u$ grows to a large negative value, $\tilde v$ grows
to a large positive value and the effective action (\ref{eq:replica})
becomes thermodynamically unstable. 
This behavior is usually interpreted as a signature
of a fluctuation-driven first-order phase transition. 
We illustrate this behavior
by showing the flow diagram in the non-random case with ${\tilde v}=0$
in Fig.~1.  

In case of the logarithmic interaction in two dimensions, we have only
the $k^2$ term in the gauge field propagator. In this case, the effective charge
parameter $\tilde w$ has a stable fixed point at ${\tilde w}^*=12\epsilon/(m+4)$.
Let us consider the non-random case with $\tilde v=0$ first. Then it is easy
to show that $\tilde u$ possesses a stable fixed point at
${\tilde u}^*=(m+10+\sqrt{m^2-52m-188})\epsilon/4(m+2)^2$ only when $m> m_c=
26+12\sqrt{6}\approx 55.394$. 
When $m$ is smaller than $m_c$, all points flow to large 
negative $\tilde u$ values as shown in Fig.~2. 
We cautiously interpret this as a signature of a first-order phase transition
\cite{fot}.
In the random case, we substitute the fixed point value of $\tilde w$
into the recursion relations for $\tilde u$ and $\tilde v$ and
find a stable fixed point with positive ${\tilde u}^*$,
${\tilde v}^*$ and ${\tilde w}^*$ when $m=1$. The fixed point values are
\begin{equation}
{\tilde u}^*=\frac{11+\sqrt{409}}{40}\epsilon,~~~
{\tilde v}^*=\frac{\sqrt{409}}{40}\epsilon,~~~
{\tilde w}^*=\frac{12}{5}\epsilon.
\end{equation}
This point has three irrelevant eigenvalues $-\epsilon$ and 
$-(33\pm\sqrt{2183}i)\epsilon/20$. 
The complex eigenvalues suggest that the renormalization-group flows spiral
into the stable fixed point and give oscillatory corrections to scaling.
In Fig.~3, we show the flow diagram in the $(\tilde u, \tilde v)$ plane
with $\tilde w$ fixed at ${\tilde w}^*$. There is a separatrix dividing the 
domain of attraction to the random fixed point and the unstable region.
The critical exponents associated with the new random fixed point are
\begin{eqnarray}
&&\eta=\frac{2}{5}\epsilon,~~~
z=1-\frac{4}{5}\epsilon,~~~
\frac{1}{\nu}=2-\left(\frac{9}{10}+\frac{\sqrt{409}}{20}\right)\epsilon,
\nonumber\\
&&\frac{1}{\nu_\tau}=\frac{1}{z\nu}=2+\left(\frac{7}{10}-\frac{\sqrt{409}}{20}
\right)\epsilon,
\end{eqnarray}
where $\nu$ and $\nu_\tau$ are the correlation length and time exponents
and $\eta$ is defined by $2\zeta_\psi+d+z=2-\eta$.

In conclusion, we performed a renormalization-group analysis 
of the superfluid-insulator transition in a particle-hole symmetric model 
of long-range interacting bosons with or without
a time-dependent random potential. 
We find a new stable fixed point with non-zero disorder and
long-range interaction when the interaction is logarithmic. Without a random
potential, the transition in logarithmically interacting systems 
is argued to be first-order.
We propose a possibility that our model may be relevant in interpreting
the vortex liquid-vortex glass transition of interacting vortex lines 
in type-II superconductors.

This work has been supported by the Korea Science and Engineering Foundation
through grant number 961-0209-052-2.

\begin{figure}
\protect\centerline{\epsfxsize=5in \epsfbox{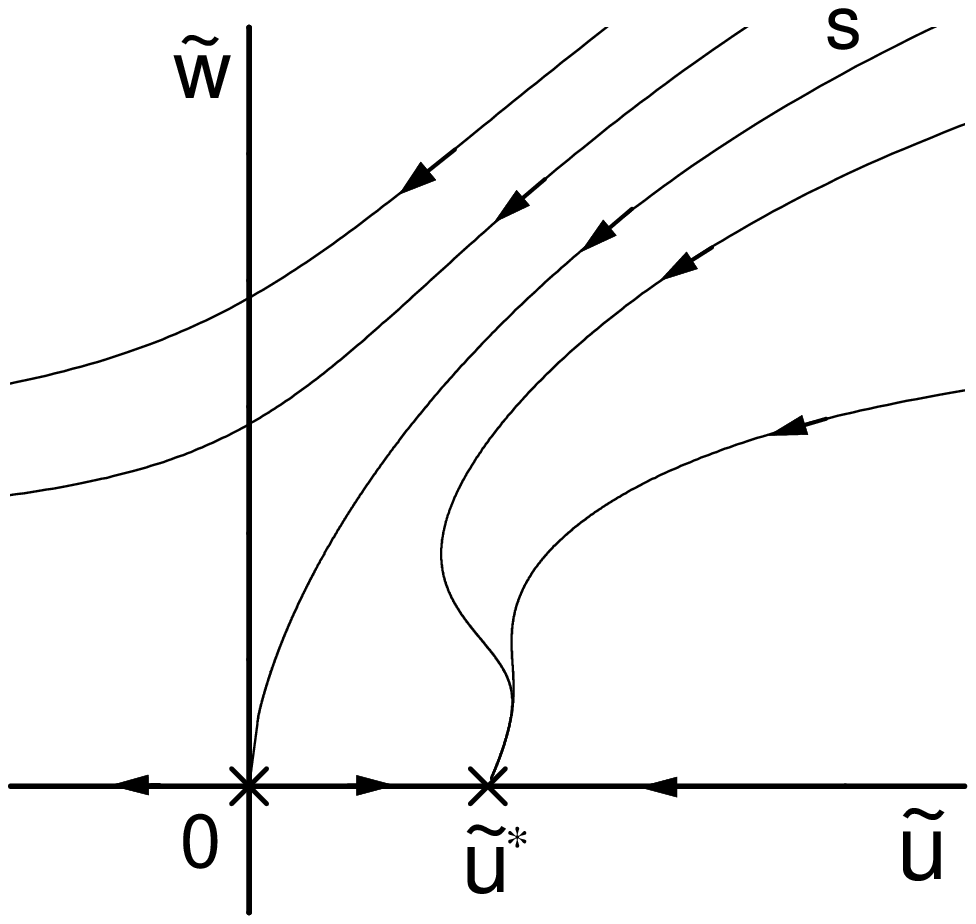}}
\caption{Flow diagram for the non-random Bose system 
interacting by the $1/r$ Coulomb potential when $m=1$.
{\it S} labels the separatrix dividing the domain of attraction to 
the XY fixed point and the unstable region. }
\end{figure}
\begin{figure}
\protect\centerline{\epsfxsize=5in \epsfbox{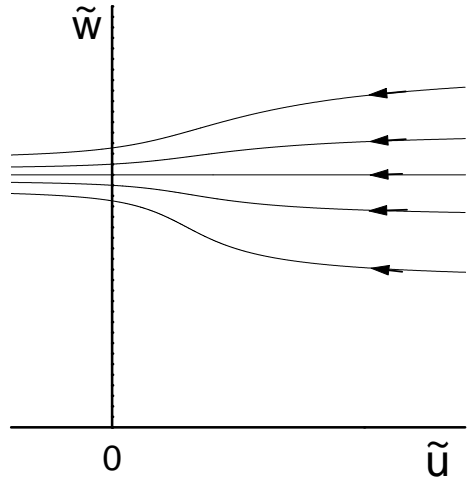}}
\caption{Flow diagram for the logarithmically-interacting
non-random Bose system when $m=1$.}
\end{figure}
\begin{figure}
\protect\centerline{\epsfxsize=5in \epsfbox{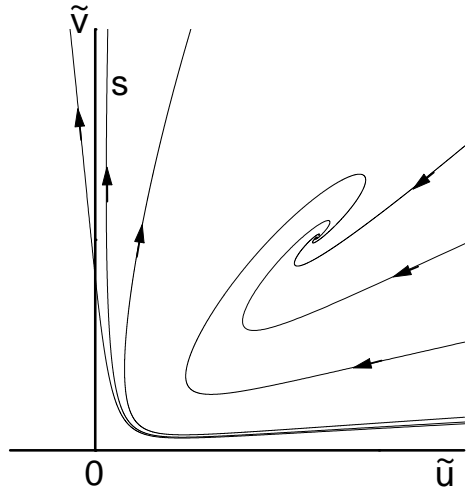}}
\caption{Flow diagram for the logarithmically-interacting
Bose system in a time-dependent random potential when $m=1$
and $\tilde w=\tilde w^*=12\epsilon /5$.
{\it S} labels the separatrix dividing the domain of attraction
to the random fixed point with positive $\tilde u$, $\tilde v$
and $\tilde w$ values and the unstable region.}
\end{figure}

\end{document}